\documentclass[twocolumn,showpacs,preprintnumbers,amsmath,amssymb]{revtex4}
\usepackage{amsfonts}
\usepackage{dsfont}
\usepackage{amsxtra}
\usepackage{hyperref}

\usepackage{graphicx,epsfig}
\usepackage{dcolumn}
\usepackage{bm}

\newcommand{\be}{\begin{equation}}
\newcommand{\ee}{\end{equation}}
\newcommand{\bea}{\begin{eqnarray}}
\newcommand{\eea}{\end{eqnarray}}
\newcommand{\beaa}{\begin{eqnarray*}}
\newcommand{\eeaa}{\end{eqnarray*}}
\newcommand{\ba}{\begin{array}}
\newcommand{\ea}{\end{array}}
\newcommand{\bi}{\begin{itemize}}
\newcommand{\ei}{\end{itemize}}
\newcommand{\ben}{\begin{enumerate}}
\newcommand{\een}{\end{enumerate}}

\newcommand{\bra}{\langle}
\newcommand{\ket}{\rangle}

\newcommand{\lb}{\label}
\newcommand{\g}{\gamma}

\newcommand{\al}{\alpha}
\newcommand{\bt}{\beta}
\newcommand{\p}{\partial}
\newcommand{\dl}{\delta}
\newcommand{\Dl}{\Delta}

\newcommand{\vp}{\varphi}

\newcommand{\Om}{\Omega}

\newcommand{\sm}{\sigma}

\newcommand{\lmax}{15}
\newcommand{\nphrnd}{10^4}
\newcommand{\ntests}{100}
\newcommand{\dmFracRND}{0.04}

\newcommand{\smallScale}{0.42}

\newcommand{\Fermi}{\textsl{Fermi} }

\begin{document}

\title{Spherical harmonics analysis of \Fermi gamma-ray data\\
and the Galactic dark matter halo}

\author{Dmitry Malyshev}
 \email{dm137 at nyu.edu}
 \altaffiliation{On leave of absence from ITEP, 
 B. Cheremushkinskaya 25, Moscow, Russia} 
\author{Jo Bovy}
 \email{jb2777 at nyu.edu}

\author{Ilias Cholis}
 \altaffiliation{
 Astrophysics Sector,
 La Scuola Internazionale Superiore di Studi Avanzati and
 Istituto Nazionale di Fisica Nucleare, Sezione di Trieste,
 via Bonomea 265, 34136 Trieste, Italy }
 \email{ilias.cholis at sissa.it}
 \affiliation{%
Center for Cosmology and Particle Physics\\
 4 Washington Place, Meyer Hall of Physics, NYU, New York, NY 10003
}%

\date{\today}

\begin{abstract}
\noindent 
We argue that the decomposition of gamma-ray maps in spherical harmonics is a sensitive tool to study dark matter
(DM) annihilation or decay in the main Galactic halo of the Milky Way.
Using the spherical harmonic decomposition in a window excluding the Galactic plane,
we show for one year of \Fermi data
that adding a spherical template (such as a line-of-sight DM annihilation profile)
to an astrophysical background significantly reduces $\chi^2$ of the fit to the data.
In some energy bins the significance of this DM-like fraction is above three sigma.
This can be viewed as a hint of DM annihilation signal, although astrophysical 
sources cannot be ruled out at this moment.
We use the derived DM fraction as a conservative upper limit on
DM annihilation signal.
In the case of $b\bar{b}$ annihilation channel the limits are 
about a factor of two less constraining than the limits from dwarf galaxies.
The uncertainty of our method is dominated by systematics related to modeling the astrophysical background.
We show that with one year of \Fermi data 
the statistical sensitivity would be sufficient to detect DM annihilation 
with thermal freeze out cross section for masses below 100 GeV.

\end{abstract}


\maketitle




 \section{Motivation}

Out of all indirect searches for dark matter (DM),
gamma-rays are probably the most ``direct"
\citep{Zeldovich:1980st, 2008Natur.456...73S}.
Charged particles, such as positrons and antiprotons,
are deflected in the Galactic magnetic field.
The information about their source is lost and only
anomalies in the spectrum may signal the presence of DM.
Most of gamma-rays propagate freely inside the Galaxy and,
together with the spectrum, they carry information about the morphology of the source.
This property may be crucial in separating a DM
signal from astrophysical backgrounds, e.g., \cite{2009PhRvL.102x1301S}.

From cosmological simulations
\cite{Navarro:1996gj, 2008Natur.454..735D, 2008ApJ...686..262K,
2008MNRAS.391.1685S}, we expect that cold dark matter in our Galaxy has
formed a nearly spherical halo with density growing toward the
Galactic center (GC).  Thus, DM annihilation or decay may
be a source of gamma-rays with a spherical shape peaked at the GC, in
addition to astrophysical and extra-galactic sources.

In this paper, we study the contribution from the main spherical
halo ignoring DM substructure.  
In order to minimize the astrophysical flux,
we mask the Galactic plane and resolved gamma-ray point sources.
The problem is that at high latitudes
a possible DM annihilation signal is relatively smooth and most probably 
subdominant to Galactic and extragalactic diffuse emission.
In the paper we propose to use the spherical harmonics decomposition of gamma-ray data
to search for DM annihilation or decay.
The contribution of a smooth signal with small amplitude 
is maximal for spherical harmonics with small angular numbers $\ell$.
Consequently, the Galactic DM signal away from the GC
may contribute most significantly to small $\ell$ harmonics,
while its contribution to large $\ell$ harmonics can be neglected
compared to the Poisson noise.

Spherical harmonics decomposition has several advantages compared to template fitting
in coordinate space:
\vspace{-2mm}
\begin{enumerate}
\item
{\bf Organization of data:}
small $\ell$ harmonics carry all the information about the large-scale distribution of sources,
while large $\ell$ harmonics are dominated by the Poisson noise.
Spherical harmonics decomposition is a linear transformation that has no information
loss, but only relevant information for large-scale distributions
is used in fitting.
\item
{\bf Universality:} small $\ell$ harmonics are insensitive to the resolution of 
pixel maps (for sufficiently small pixel sizes).
In particular, the templates may have different resolution and do not need to be
brought to the same pixel size as in the case of coordinate space fitting.
The $\chi^2$ is also independent of resolution,
while in coordinate space the absolute likelihood depends on the number of pixels.
\item
{\bf Linearity and stability:}
both the transformation of data from coordinate space
and the fitting in spherical harmonics space are linear operations
(the $\chi^2$ has the usual quadratic form).
Thus, a nonlinear Poisson likelihood in coordinate space 
is substituted by a combination of two linear operations 
in spherical harmonics space.
This may be useful for stability of the fitting procedure in the case of small 
numbers of photons in pixels:
the Poisson probability is undefined for nonpositive expected numbers of photons,
while small negative expected numbers should not a be problem:
it simply means that the template is not perfect and
we over subtract this template to fit data somewhere else.
\item 
{\bf Symmetry:}
in some cases spherical harmonics may be useful
to focus on a part of data with a particular symmetry in mind.
In this paper we will use the spherical symmetry
of dark matter distribution around the Galactic center.
If we point the $z$-axis toward the GC, then
dark matter contributes only to $Y_{lm}$ harmonics with $m = 0$,
i.e., we may select only $a_{l0}$ modes in fitting.
\end{enumerate}

A computational algorithm for fitting in spherical harmonics space is
straightforward but there are a few things to keep in mind.  First,
the $Y_{lm}$'s are not orthogonal in a window on the sphere.  The
corresponding spherical modes, $a_{lm}$'s, are still independent but
correlated. As a result, their covariance matrix is not diagonal.  In
general, this may render the computations unfeasible, unless one uses
some special techniques (such as the Gabor transform for the
power spectrum \citep{2002MNRAS.336.1304H}).  We will use only a few
$Y_{l0}$ harmonics corresponding to the largest scales.  The
corresponding covariance matrix is relatively small and can be easily
computed.

The choice of the astrophysical background model is a more conceptual
problem.  A thorough solution of this problem can be quite complicated
and we will not discuss it here.  The main purpose of our work will be
to illustrate the method of the spherical harmonics transform in the analysis of
gamma-rays.  As a toy model for the astrophysical background we will
use the gamma-ray distribution in a low-energy bin, since we expect
that the DM contribution to the spectrum is insignificant at low
energies.

The paper is organized as follows.
In Section \ref{sect:method} we describe an algorithm
of fitting templates in spherical harmonics space.
In Section \ref{sect:data_analysis} we apply this method
for \Fermi gamma-ray data.
We compare two cases.
In the first case we use two templates: a low-energy bin and an isotropic 
distribution.
In the second case we also add a distribution of photons with a 
spherical symmetry around the GC.
We find that the residual for the three-template models
has a much better $\chi^2$ than the residual for the two-template model.
In Section \ref{sect:global_model} we find 
the best-fit energy spectrum of the fluxes assuming a power-law 
dependence on energy.
We also put constraints on DM annihilation in $b\bar{b}$.
Section \ref{sect:concl} has conclusions.

There are three appendixes.
In Appendix \ref{sect:shcov} we calculate the covariance matrix for spherical
harmonics defined on a window in the sphere.
In Appendix \ref{sect:mct} we check the fitting algorithm with a Monte Carlo simulation.
In Appendix \ref{sect:ps} we discuss the contribution to the angular power spectrum from point sources.

\section{\lb{sect:method} Method}
\lb{sect:fitting}

In this section we describe a general method of template fitting in 
spherical harmonics space.
In the next section we apply this method to the \Fermi gamma-ray data
to search for DM annihilation in the Milky Way halo.

In general, an algorithm will contain several steps:
\begin{enumerate}
\item Choose a mask (for instance, one can mask the Galactic plane and point sources).
\item Find the spherical harmonics decomposition of the data outside of the masked region, $a_{lm}$.
\item Find the covariance matrix for the spherical harmonics coefficients, ${\rm Cov}(a_{lm}, a_{l'm'})$.
\item Formulate a model for the gamma-ray distribution as a function of parameters $\al$
and find the corresponding decomposition into spherical modes $b_{lm}(\al)$ outside of the mask.
\item Find the best-fit model parameters $\al_*$ by minimizing a $\chi^2$,
where we use the full covariance matrix instead of $\sm^2$
due to a nontrivial correlation of the spherical modes on a window in the sphere
(Equation (\ref{eq:chi2_ini})).
\end{enumerate}

In the remaining part we will mostly introduce the notations that will
be necessary in interpreting the results of data analysis in the next section.
The mathematical details can be found in Appendix \ref{sect:shcov}.

In the calculations of spherical harmonics decomposition,
it is convenient to use a pixelation of the sphere (we use HEALPix \cite{2005ApJ...622..759G}).
We will consider some energy bins $E_i$ and 
denote by $n_p(E_i)$ the number of photons 
inside the energy bin $i$ in a pixel $p = 1,\ldots,N_{\rm pix}$.
For clarity, in the following formulas we will suppress the dependence on energy.

Define the photon density as $\rho(\g_p) = \frac{n_p}{\dl \Om_p}$,
where $\dl \Om_p$ is the size 
and $\g_p$ is the center of the pixel $p$.
We put $n_p = \rho(\g_p) = 0$, if the center of the pixel is inside the mask.
The spherical harmonics transform of the gamma-ray data is
\be
\lb{eq:alms}
a_{lm} = \int Y^*_{lm}(\g) \rho(\g) d\Om
\approx \sum_{p = 1}^{N_{\rm pix}} Y^*_{lm}(\g_p) n_p.
\ee
In Appendix \ref{sect:shcov} we show that the covariance matrix
has the following simple expression
\be
\lb{eq:var0}
{\rm Cov}(a_{lm}, a_{l'm'}) = \sum_p Y^*_{l'm'}(\g_p) Y_{lm}(\g_p) n_p\,.
\ee

Let us now describe the parametrization of the models.
In general, the shape of the model fluxes can depend on some parameters.
In this paper we will focus on the template fitting,
where the shape of the templates for various sources is fixed
and the only variable parameters are the normalizations
in every energy bin.

For each template $\al$ in every energy bin,
the variable parameter will be the average flux inside the window
$F^\al$ in units of $\frac{1}{\rm GeV\: cm^2\: s\: sr}$.
In order to find the number of photons in a pixel $p$ from a template 
$\al$, one needs to multiply $F^\al$ by the \Fermi exposure $E_p$
times the probability distribution function (PDF) $\rho_p^\al$ that carries the information
about the shape of the template $\al$, times the size of the window $\Om_{\rm W}$,
times the size of the energy bin $\Dl E$.
The number of photons from the template $\al$ in a pixel $p$ is
\be
n_p^\al = F^\al \cdot (E_p \cdot \rho_p^\al \cdot \Om_{\rm W} \cdot \Dl E),
\ee
where the PDF is normalized as $\sum_p \rho_p^\al = 1$.
Let us denote the spherical harmonics decomposition of the function
in the parenthesis as
\be
v_{lm}^\al = \sum_{p = 1}^{N_{\rm pix}} Y^*_{lm}(\g_p) 
(E_p \cdot \rho_p^\al \cdot \Om_{\rm W} \cdot \Dl E),
\ee
then the spherical harmonics decomposition of 
a linear sum of sources is
\be
b_{lm}(F^\al) = \sum_\al F^\al v_{lm}^\al 
\ee
The best-fit fluxes $F^\al_*$ can be found from minimizing the $\chi^2$
in Equation (\ref{eq:chi2_ini}).

The uncertainty of the model parameters $F^\al$ around the minimum 
of $\chi^2$ can be estimated from the Hessian matrix
\be
H_{\al \bt} = \left.\frac{\p^2 \chi^2}{\p F^\al \p F^\bt}\right|_{F^\al = F^\al_*}.
\ee
In particular, the variance of the model fluxes 
can be estimated as
\be
\lb{eq:var_f}
{\rm Var} (F^\al) = {H^{-1}}_{\al \al}.
\ee

\section{Data analysis}
\lb{sect:data_analysis}

\subsection{Data selection}
\lb{sect:data}

\begin{figure*}[tb] 
\begin{center}
\hspace{-3mm}
\epsfig{figure = 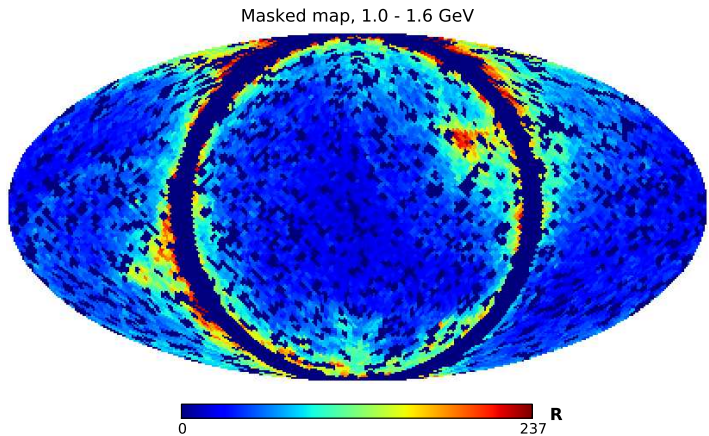,scale=0.45}
\end{center}
\vspace{-8mm}
\noindent
\caption{\small 
Counts in pixels for 1 - 1.6 GeV energy bin.
The Galactic center is in the north pole, the anti-center is in the south pole
(southern hemisphere is in the center).
We mask $|b| < 10^\circ$ and gamma-ray point sources
detected by \textsl{Fermi} \citep{2010ApJS..188..405A}.
}
\label{fig:data}
\vspace{1mm}
\end{figure*}

We consider 13 months of \Fermi gamma-ray data 
(August 4, 2008 to August 25, 2009)
that belong to the ``diffuse class'' (Class 3) of the LAT pipeline. 
We exclude the data beyond zenith angles of $105^{\circ}$ due to significant contamination 
from atmospheric gamma-rays.
We also exclude the data taken over the South Atlantic Anomaly (SAA)
and mask the point sources detected by \Fermi \citep{2010ApJS..188..405A}.

Most of the time we will use the gamma-rays between 1 GeV and 300 GeV
which we separate in 10 exponential energy bins
between 1 GeV and 100 GeV plus an extra energy bin between 100 GeV and 300 GeV.
We mask the pixels centered within $10^\circ$
from the Galactic plane.
We also mask all pixels that either contain a gamma-ray point source
or if the boundary of the pixel 
is within 68\% containment angle
$\approx 0.7^\circ$ at $E = 1$ GeV 
\cite{2009ApJ...697.1071A}
to a gamma-ray point source.

The interpretation of spherical harmonics decomposition of a DM
model is simplest in the coordinate system where the $z$-axis points
toward the GC (at odds with the standard Galactic coordinates in
which the $z$-axis points toward the Galactic North pole).  We
choose the $x$-axis pointing toward the Galactic South pole.  
If we were considering all data without masking,
then DM would contribute only to $Y_{l0}$ harmonics due to a rotational symmetry 
round the new $z$-axis.
In the presence of a mask, DM contributes to all spherical harmonics, but
its contribution to $Y_{l0}$ modes is still maximal and we will restrict our attention
to these modes for simplicity of analysis.

Pixelation of the data and spherical harmonics decomposition is performed with {\it healpy}, 
the python version of HEALPix \citep{2005ApJ...622..759G}%
\footnote{\url{http://code.google.com/p/healpy/}, \url{http://HEALPix.jpl.nasa.gov}.}.

An example of a gamma-ray counts map for the energy bin between 1 GeV and 1.6 GeV
with the $z$-axis pointing toward the GC can be found in Figure \ref{fig:data}.

Summary of data selection and model parameters:
\ben
\item
Mask the gamma-ray point sources
and the Galactic plane within  $|b| < 10^\circ$.
\item
Rotate the $z$-axis to point toward the GC.
\item
Consider $Y_{lm}$ harmonics with $\ell \leq 15$ and $m = 0$.
\item 
We choose HEALPix parameter nside = 32 
(corresponding to pixel size of about $2^\circ$).
\item
Astrophysics template energy bin: 1 GeV $ < E < $ 1.6 GeV.
\een
These are the data selection parameters for the ``main'' example that we consider
in Section \ref{sect:bin2bin}.
In Section \ref{sect:vars} we consider the effect of varying these parameters.

\subsection{Bin to bin fitting}
\lb{sect:bin2bin}

\begin{figure*}[th] 
\begin{center}
\hspace{-3mm}
\epsfig{figure = 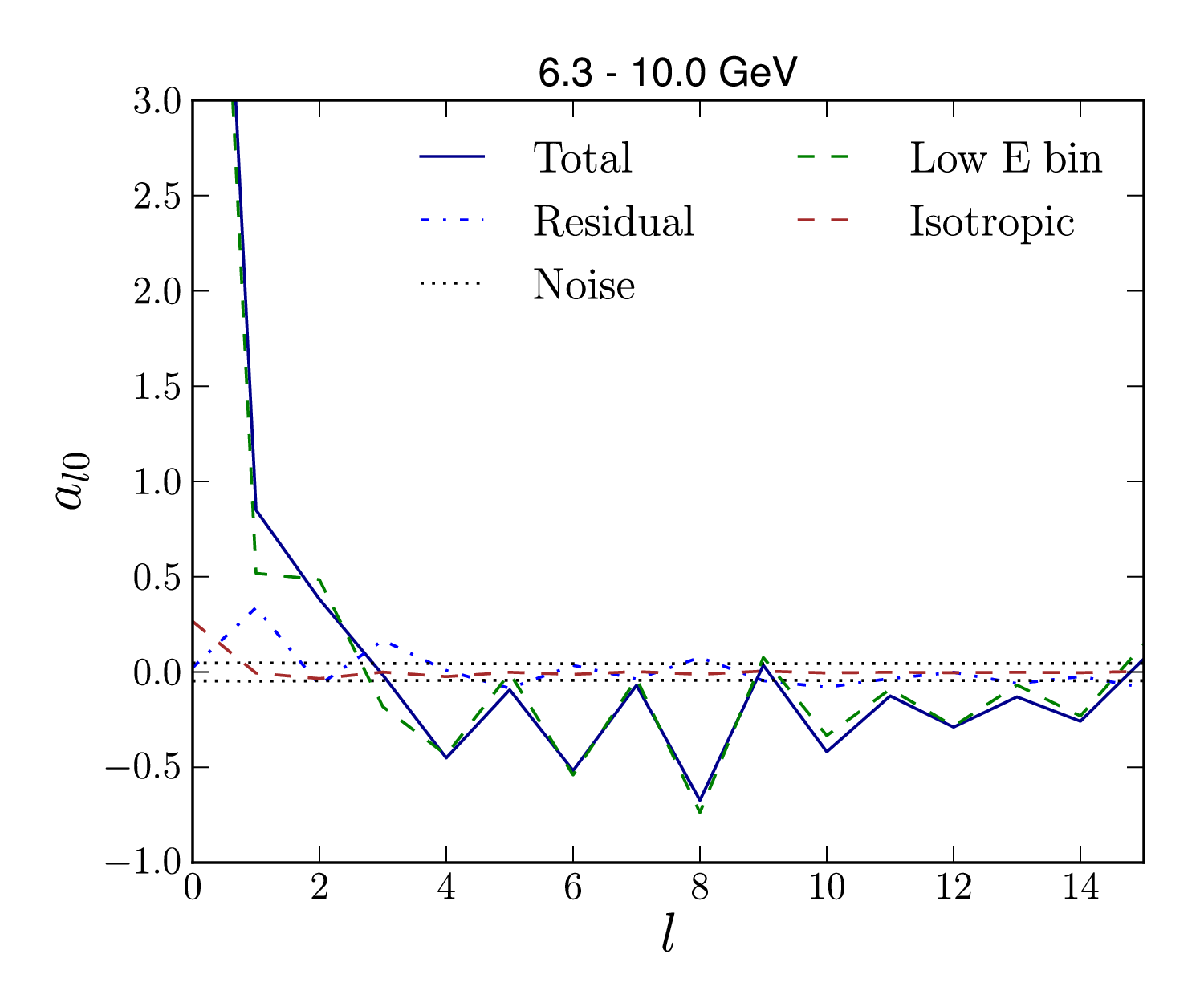,scale=0.55}
\epsfig{figure = 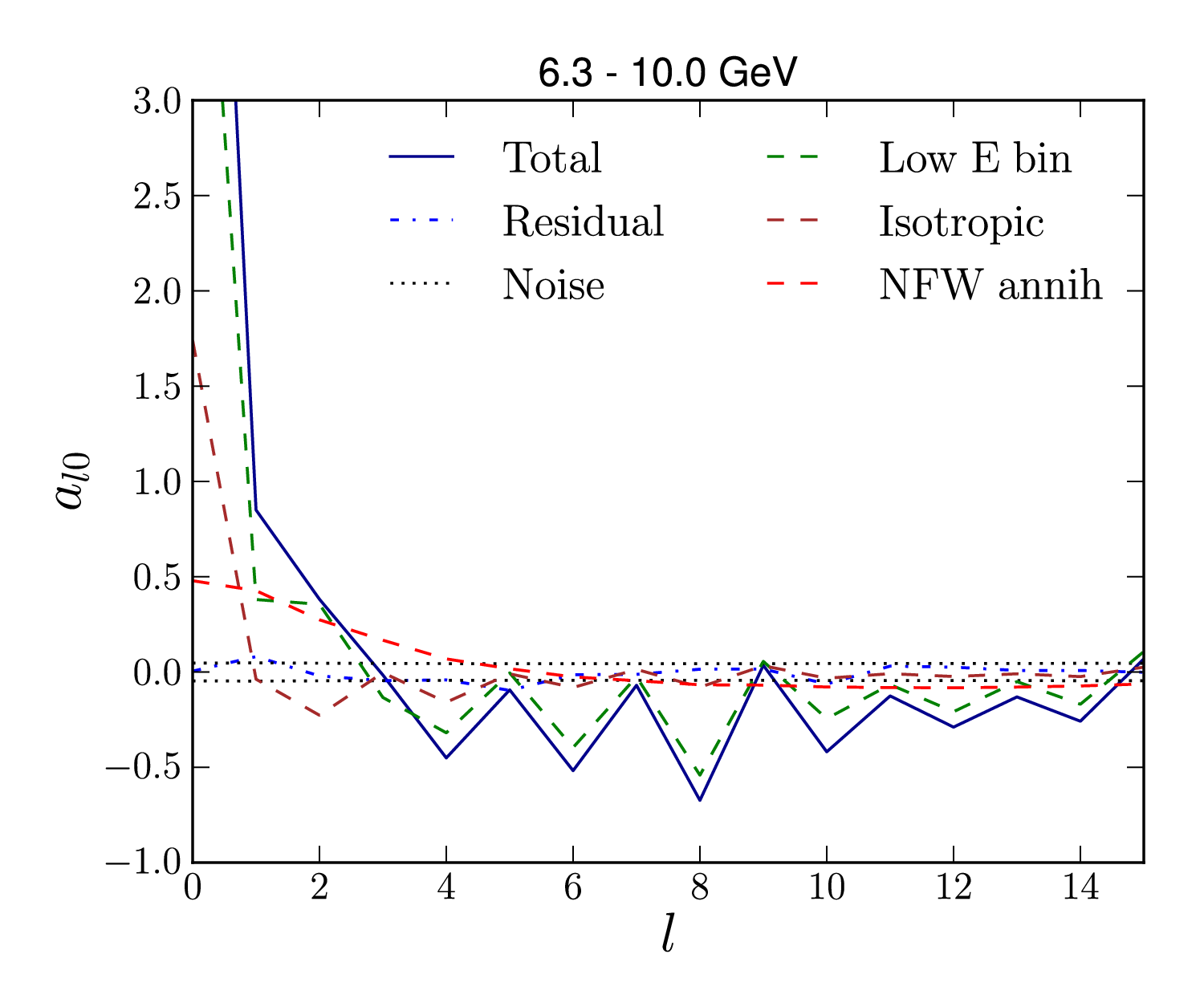,scale=0.55}
\end{center}
\vspace{-8mm}
\noindent
\caption{\small 
An example of the $a_{lm}$ fitting procedure for $m = 0$ and $\ell = 0,\ldots, 15$
in the 6.3 GeV to 10 GeV energy bin.
The harmonic decomposition in a window is defined in Section \ref{sect:fitting}
(there is an additional HEALPix normalization factor $4\pi/N_{\rm pix}$
with respect to Equation (\ref{eq:alms})).
The low-energy bin  $a_{lm}$'s are given by the gamma-ray data in 1 GeV to 1.6 GeV energy bin.
The isotropic template $a_{lm}$'s are nonzero for $\ell > 0$ due to the spherical harmonics decomposition in a
window.
Large fluctuations of data $a_{lm}$'s are mostly due to symmetry of the mask
and the fact that the Galactic flux is stronger near the Galactic plane
(notice, that the modes with even $\ell$ are significantly larger than the modes with odd $\ell$'s).
The noise level is given by the square root of the diagonal elements in the
covariance matrix in Equation (\ref{eq:var0}).
The $\chi^2$ for the two-template fit on the left is 80 for 14 dof,
while the three-template fit on the right has $\chi^2 = 16$ for 13 dof. 
The $\chi^2$ in other energy bins can be found in Section \ref{sect:profiles},
where we also compare the NFW annihilation profile with other profiles.
}
\label{fig:alms}
\vspace{1mm}
\end{figure*}

We use the energy bin between 1 GeV and 1.6 GeV as a template for the
Galactic astrophysical emission
assuming that the gamma-ray emission at these energies
is dominated by the Galactic cosmic ray production
through the $\pi^0$ decay.
One of the possible limitations of this template is the 
inverse Compton scattering (ICS) component of gamma-rays.
Relative contribution from the ICS photons increases with energy and may become comparable
to $\pi^0$ photons above 10 GeV \cite{2010PhRvL.104j1101A}.
As a result, a low-energy bin template may underestimate the ICS component at higher energies.
Currently, there is no generally accepted model for the ICS emission
(see, e.g., the caveats section in the description of the \Fermi diffuse background
\href{http://fermi.gsfc.nasa.gov/ssc/data/access/lat/ring_for_FSSC_final4.pdf}{model}
\footnote{The \Fermi background emission model can be found at
\url{http://fermi.gsfc.nasa.gov/ssc/data/access/lat/}}).
We will treat the ICS component as a systematic uncertainty in the current analysis.

We consider an isotropic template as a model for the extragalactic emission.
We will also consider several templates with a spherical distribution around the GC.
Our main working example will be the line-of-sight DM annihilation
in Navarro, Frenk, and White (NFW) profile \citep{Navarro:1996gj}.
The NFW profile is
\be
\lb{eq:NFW}
\rho_{\rm DM}(r) \propto \frac{r}{r_s}\frac{1}{(1 + r / r_s)^2}
\ee
where the scale parameter $r_s = 20$ kpc
\cite{2007MNRAS.379..755S,2008ApJ...684.1143X}.
The window angle $|b| > 10^\circ$ corresponds to distances $r > 1.5$ kpc from the GC.
At these distances the NFW profile is similar to less cuspy profiles,
such as the Einasto profile \citep{Navarro:2003ew}.
The annihilation signal is proportional to $\rho_{\rm DM}^2$,
cf., Equation \ref{eq:DMline}.

\begin{figure*}[th] 
\begin{center}
\hspace{-3mm}
\epsfig{figure = 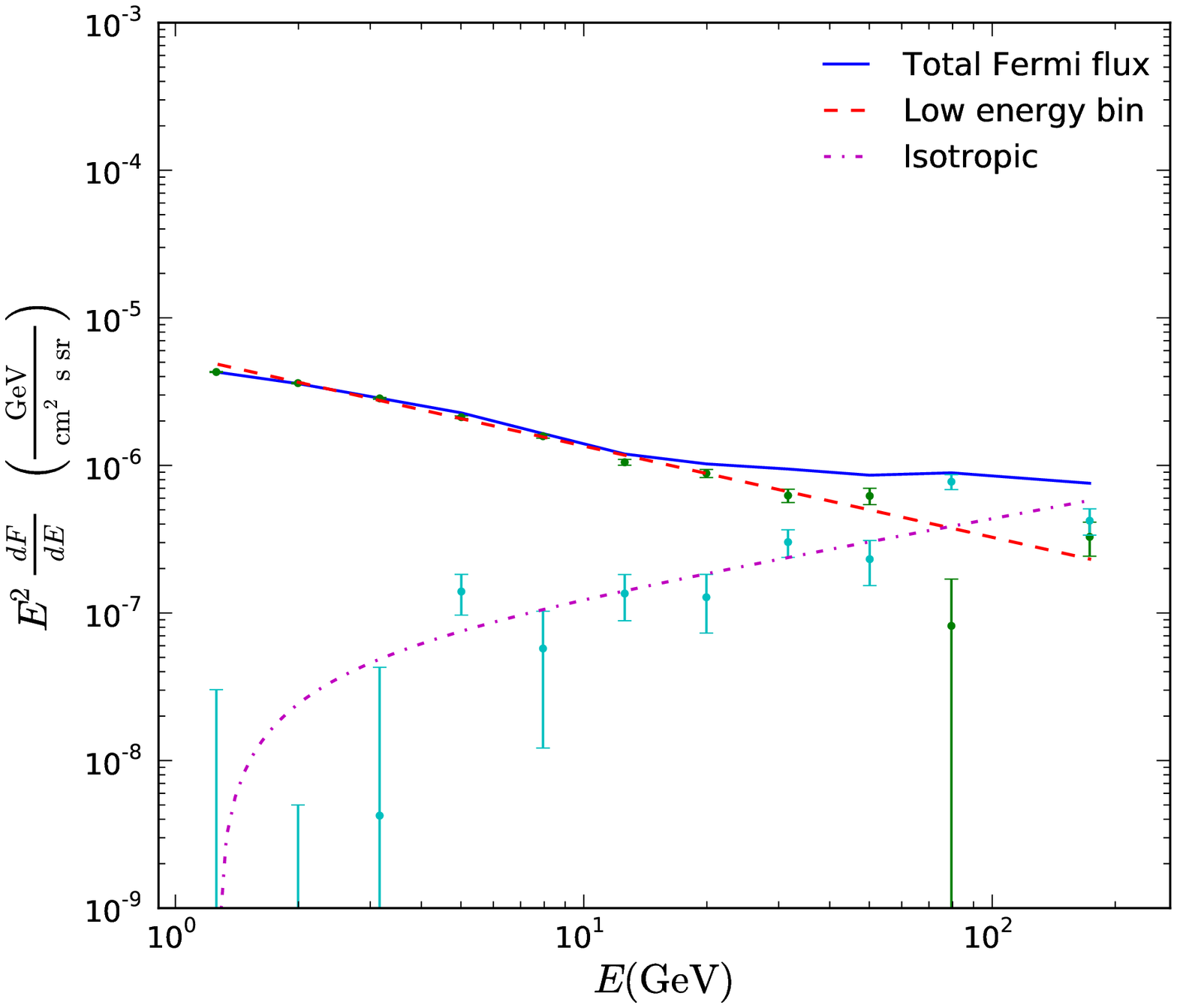,scale=0.42}
\epsfig{figure = 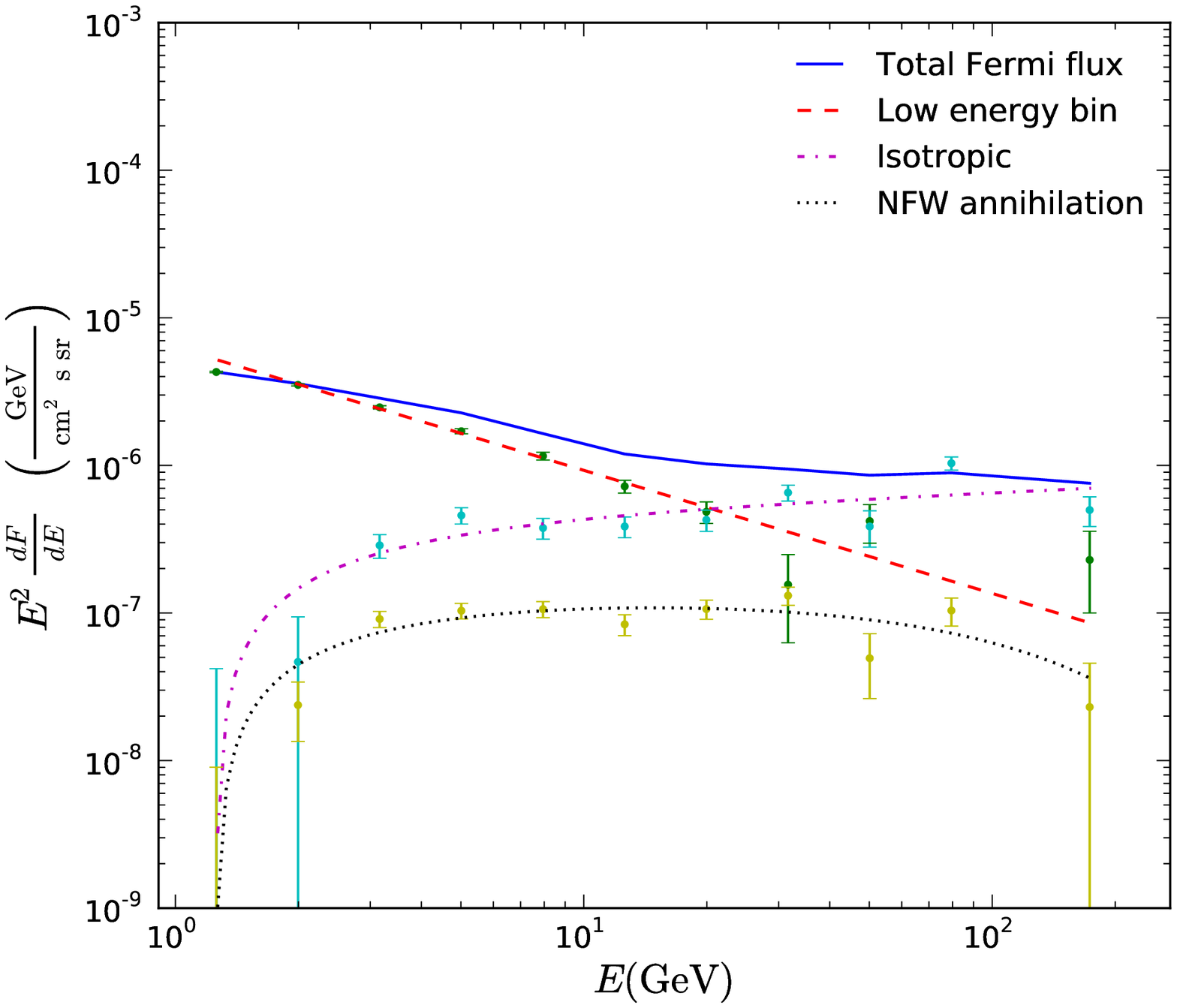,scale=0.42}
\end{center}
\vspace{-8mm}
\noindent
\caption{\small 
Left plot: two-template model fitting.
Right plot: three-template model fitting.
Points represent fitting in every energy bin independently.
Lines represent best-fit with power-law spectra
discussed later in Section \ref{sect:global_model}.
The first bin is ``singular" since we use the data in this bin as one of the templates.
The points associated with the NFW annihilation template are significantly above zero,
i.e., there is a significant signal that grows toward the GC,
but, as we show in Figure \ref{fig:chi2},
other profiles with rotational symmetry around the GC
give similar improvement of $\chi^2$, i.e., 
current data are not sufficient to distinguish different profiles
and understand the nature of the signal.
In Section \ref{sect:dm_lim} we will use the points associated with the NFW annihilation
profile to put conservative upper limits on annihilation cross-section.
}
\label{fig:astroDM}
\vspace{1mm}
\end{figure*}

As a null hypothesis we take a combination of two templates:
an astrophysical template modeled by a low-energy bin and an isotropic flux.
We compare this model with the three-template model, where we also add
a template with spherical symmetry around the GC.
In fitting, we decompose the templates 
and the data in every energy bin into spherical harmonics
according to the algorithm in Section \ref{sect:fitting}.
The fluxes corresponding to the templates are found
by minimizing the $\chi^2$ in Equation (\ref{eq:chi2_ini}).
An example of fitting the data in an energy bin between 6.3 and 10 GeV
using the spherical harmonics is presented in Figure \ref{fig:alms}.
The results of fitting in every energy bin are
shown in Figure \ref{fig:astroDM}.

The best-fit value of the flux corresponding to the spherical template
can be used to put a conservative upper bound on DM annihilation
into a pair of monochromatic photons (the DM line signal).
The upper bounds obtained this way are an order of magnitude less constraining than the 
limits on monochromatic gamma-rays obtained by the \Fermi Collaboration
\citep{2010PhRvL.104i1302A}.
The spherical harmonics decomposition method has a better performance
for signals with smooth energy dependence.
In Section \ref{sect:dm_lim} we find conservative limits on $b\bar{b}$ DM annihilation
only a factor of a few less constraining than 
the limits from dwarf spheroidal galaxies \cite{2010ApJ...712..147A}.

\subsection{\lb{sect:vars} Variation of parameters}

\begin{figure}[th] 
\begin{center}
\epsfig{figure = 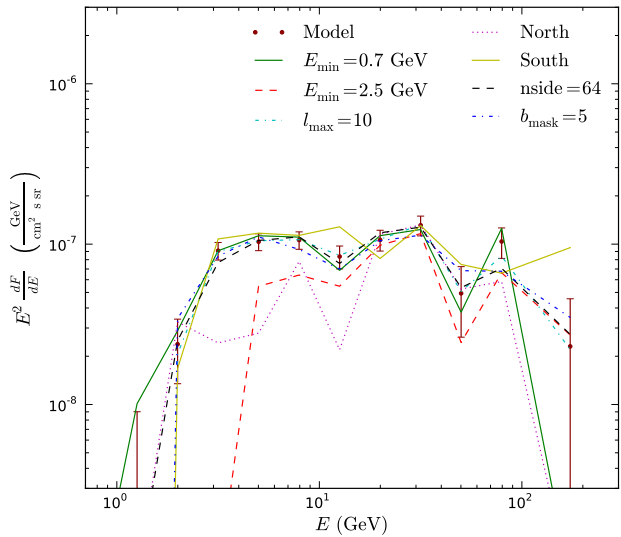,scale=0.45}
\end{center}
\vspace{-8mm}
\noindent
\caption{\small 
Dependence of flux associated with the NFW annihilation template on model parameters.
The main model and the variations are described in Sections \ref{sect:bin2bin}  and \ref{sect:vars}.
}
\label{fig:vars}
\vspace{1mm}
\end{figure}

In this subsection we 
check the robustness of spherical harmonics decomposition
with respect to variation of data selection and model parameters.
The main model is given by the flux associated with
NFW annihilation template found in Section \ref{sect:bin2bin}
with parameters defined in Section \ref{sect:data}.
We consider the following variations of parameters (Figure \ref{fig:vars}):
\begin{enumerate}
\item
Different energy bins for the astrophysics template:
0.7 GeV $< E <$ 1 GeV and  2.5 GeV $< E <$ 4 GeV.
The contribution above 10 GeV is consistent with the 
main model.
\item 
The DM flux does not change significantly if we
shrink the window size to $|b| < 5^\circ$,
change the number of harmonics to $\ell_{\rm max} = 10$,
or change the resolution to nside = 64 (pixel size of about $1^\circ$).
\item
Separating the northern and the southern hemispheres: 
the contribution of the spherical template in
the north is significantly smaller than in the south.
This may be due to a stronger astrophysical flux in the north
(compare with the Schlegel, Finkbeiner, Davis (SFD) dust template 
\cite{Schlegel:1997yv}
that is believed to trace
the $\pi^0$ production of gamma-rays 
\cite{2010ApJ...717..825D, 2010ApJ...724.1044S}).
\end{enumerate}
We performed other variations, such as increasing the window size
and increasing $\ell_{\rm max}$ (not shown in Figure \ref{fig:vars}).
We find that the variation of parameters does not change the results
significantly.

\subsection{Different spherical profiles}
\lb{sect:profiles}

\begin{figure}[th] 
\begin{center}
\epsfig{figure = 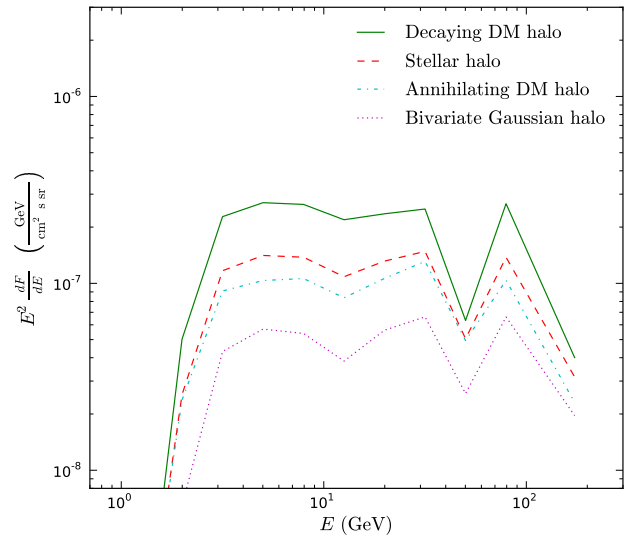,scale=0.45}
\end{center}
\vspace{-8mm}
\noindent
\caption{\small 
Best-fit values of the flux associated with different spherical profiles.
The values are found by fitting a three-template model
(low-energy bin, isotropic, and spherical templates)
to the data.
The case of NFW annihilation is discussed in detail in Section \ref{sect:bin2bin}.
}
\label{fig:profiles}
\vspace{1mm}
\end{figure}

\begin{figure}[th] 
\begin{center}
\epsfig{figure = 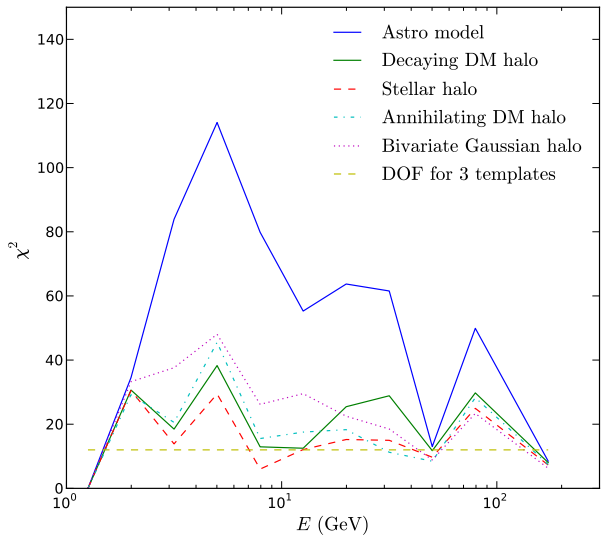,scale=0.45}
\end{center}
\vspace{-8mm}
\noindent
\caption{\small 
Goodness of fit of the models presented in Figure \ref{fig:profiles}
compared to the astrophysical model.
The number of degrees of freedom (DOF) for the three-template model is 13
(sixteen $a_{l0}$ modes minus three parameters associated with
the normalization of the templates).
}
\label{fig:chi2}
\vspace{1mm}
\end{figure}

\begin{figure}[th] 
\begin{center}
\hspace{-3mm}
\epsfig{figure = 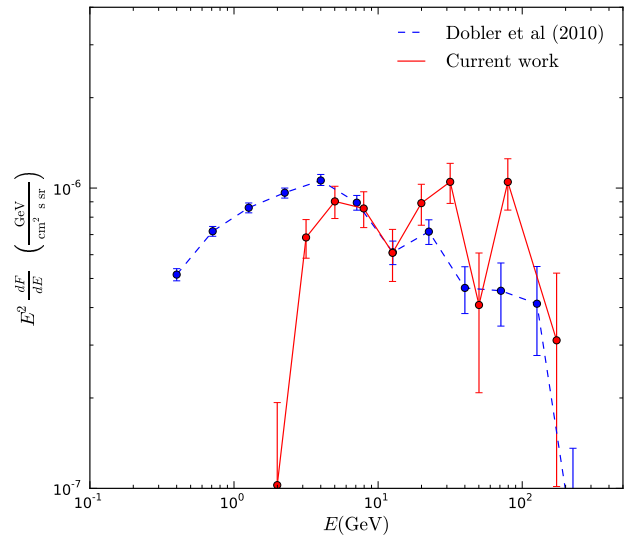,scale=.45}
\end{center}
\vspace{-8mm}
\noindent
\caption{\small 
Comparison with the gamma-ray haze spectrum in \cite{2010ApJ...717..825D}.
Our flux is the same as in Figure \ref{fig:profiles}
but averaged over the window
$l \in (-15^\circ,\: 15^\circ)$ and $b \in (-30^\circ,\: - 10^\circ)$ \cite{2010ApJ...717..825D}.
}
\label{fig:dobler}
\vspace{1mm}
\end{figure}

In this section we compare different profiles with spherical symmetry.
Together with the NFW DM annihilation profile $\propto \rho_{\rm NFW}^2$,
studied in Section \ref{sect:bin2bin},
we consider NFW DM decay $\propto \rho_{\rm NFW}$,
a profile $\propto 1/r^3$ corresponding to the distribution of mass in the stellar halo
of the Milky Way \citep{2008ApJ...680..295B, 2008ApJ...673..864J}
($r$ is the distance from the GC),
and a bivariate Gaussian profile with
$\sm_l = 15^\circ$ and 
$\sm_b = 25^\circ$
studied in \cite{2010ApJ...717..825D}.

The best-fit flux values associated with these profiles are shown in Figure \ref{fig:profiles}.
In Figure \ref{fig:chi2} we compare the $\chi^2$ for these model
with the null hypothesis (a low-energy bin template plus an isotropic flux)
from Section \ref{sect:bin2bin}.
We find that any of the spherical profiles give a significant improvement of $\chi^2$
compared to the two-template model.
The stellar halo profile has smaller $\chi^2$ than the other three profiles.
A possible astrophysical source of high energy gamma-rays at high latitudes
could be a population of millisecond pulsars 
\citep{2010JCAP...01..005F, 2010arXiv1002.0587M}.

In Figure \ref{fig:dobler} we compare our fit of the bivariate Gaussian halo with the calculation in
\cite{2010ApJ...717..825D}.
There is a general agreement above $\sim$ 2 GeV.
Our error bars are larger than the errors in
\cite{2010ApJ...717..825D},
possibly due to the usage of a subset of spherical modes
with $m = 0$ rather than all spherical harmonics.

\section{\lb{sect:global_model}  Energy spectrum}

In this section we fit the fluxes found in the previous section by
some general energy spectra.
We assume that there are three main contributions to the gamma-ray flux:
Galactic astrophysics emission, 
isotropic  emission (extragalactic plus a possible contamination from 
misidentified cosmic rays), 
and an additional spherically symmetric flux.
We fit the Galactic and the isotropic fluxes
by power-law spectra.
For the spherical template, we use a power-law with an exponential cutoff
and an energy spectrum for DM annihilating into $b\bar{b}$.

\subsection{Power-law energy spectra}

In the previous section we have used 
a low-energy bin as a template for the Galactic astrophysical component.
In reality the flux in the first bin $E_0$ is a sum of all components.
As a result there is a nontrivial relation between the fluxes associated to templates,
and the intrinsic fluxes.
Let us denote the intrinsic Galactic component of the flux by $\Phi_{\rm g}(E)$,
the intrinsic isotropic component by $\Phi_{\rm i}(E)$,
and a spherical (``dark matter") component by $\Phi_{\rm d}(E)$.
The flux associated to the low-energy bin will be denoted as $F_{\rm a}(E)$,
the flux associated to isotropic template as $F_{\rm i} (E)$,
and the flux associated to spherical templates as $F_{\rm d} (E)$.

The flux in the first energy bin is
\be
F_{\rm a}(E_0) = \Phi_{\rm g}(E_0) + \Phi_{\rm i}(E_0) + \Phi_{\rm d}(E_0)\,.
\ee
i.e., the corresponding template has contributions from the 
actual Galactic photons, from the isotropic distribution, and, possibly, 
from an additional spherical component.
Consequently, a fraction of isotropic and DM photons is included in
the flux corresponding to the ``astrophysical'' low-energy bin template at all energy bins.

\begin{table*}[tpb]
\begin{center}
\footnotesize
\begin{tabular} {|c|c|c|c|c|c|c|}
\hline
Model & $n_{\rm g}$ & $n_{\rm i}$ & $n_{\rm d}$  & $E_{\rm cut}$ (GeV) \\
\hline
Astrophysical		& 	$2.62 \pm 0.01$ & $1.49 \pm 0.03$ &  $-$ & $-$\\
Decay			& 	$2.79 \pm 0.02$ & $1.68 \pm 0.04$ & $2.13 \pm 0.05$ & 142		\\
Annihilation		&	$2.83 \pm 0.02$ & $1.88 \pm 0.03$ & $2.01 \pm 0.05$ & 137 		\\
Stellar halo		& 	$2.87 \pm 0.02$ & $1.89 \pm 0.04$ & $2.05 \pm 0.05$ & 152  	\\
Gaussian halo		&	$2.84 \pm 0.02$ & $1.86 \pm 0.04$ & $1.98 \pm 0.05$ & 184 		\\
\hline 
\end{tabular}
\end{center}
\caption{\small
Fitting the fluxes associated with different templates by power-law intrinsic
Galactic, isotropic, and spherically symmetric fluxes.
The parametrization is given in Equations (\ref{eq:astro_par}), (\ref{eq:iso_par}),
and (\ref{eq:dm_par}).
The profiles are defined in Section \ref{sect:profiles}.
}
\label{tab:vars}
\end{table*}

In this subsection, we consider the following parametrization of intrinsic fluxes
\bea
\lb{eq:flux_g}
\Phi_{\rm g} = \Phi_{\rm g0} \left(\frac{E}{ E_0}\right)^{-n_{\rm g}}\,, \\
\lb{eq:flux_i}
\Phi_{\rm i} = \Phi_{\rm i0} \left(\frac{E}{ E_0}\right)^{-n_{\rm i}}\,, \\
\lb{eq:flux_d}
\Phi_{\rm d} = \Phi_{\rm d0} \left(\frac{E}{ E_0}\right)^{-n_{\rm d}}  e^{- \frac{E - E_0}{E_{\rm cut}}}\,.
\eea
If we assume that the Galactic gamma-rays provide the most significant contribution to
the astro template at $E_0$,
then we can expect that the astro template flux $F_{\rm a}$ has the same power-law index as 
the intrinsic Galactic flux $\Phi_{\rm g}$
and the following parametrization of $F_{\rm a}$ is reasonable:
\be
\lb{eq:astro_par}
F_{\rm a} (E) = \left(\Phi_{\rm g0} + \Phi_{\rm i0} + \Phi_{\rm d0}\right) \left(\frac{E}{ E_0}\right)^{-n_{\rm g}}\,.
\ee
The fluxes for the isotropic and DM templates are equal to the intrinsic fluxes
minus the contribution to the astro template,
\bea
\lb{eq:iso_par}
F_{\rm i} (E) &=& \Phi_{\rm i0} \left(\frac{E}{ E_0}\right)^{-n_{\rm i}} 
				- \Phi_{\rm i0} \left(\frac{E}{ E_0}\right)^{-n_{\rm g}}\,, \\
\lb{eq:dm_par}
F_{\rm d} (E) &=& \Phi_{\rm d0} \left(\frac{E}{ E_0}\right)^{-n_{\rm d}} e^{- \frac{E - E_0}{E_{\rm cut}}}
				 - \Phi_{\rm d0} \left(\frac{E}{ E_0}\right)^{-n_{\rm g}}\,.
\eea

In order to find the parameters of the energy spectra,
we use the fluxes found in Section \ref{sect:bin2bin} 
in every energy bin as ``data points''.
The error bars are derived from Equations (\ref{eq:astro_par}), (\ref{eq:iso_par}),
and (\ref{eq:dm_par}).
The models of the fluxes are parameterized in Equations (\ref{eq:flux_g}),
(\ref{eq:flux_i}),(\ref{eq:flux_d}).
The best-fit indices and the cutoff are presented in Table \ref{tab:vars}.

The index $n_{\rm g} \approx 2.8$ for the Galactic component 
is consistent with the pion production of gamma-rays.
The index $n_{\rm i} \approx 1.9$ for the isotropic flux is harder than the typical indices 
$n = 2.2 - 2.5$ for extragalactic diffuse background or Active Galactic Nuclei (AGN) spectra 
\cite{2010ApJ...720..435A, 2010PhRvL.104j1101A}.
This discrepancy is most probably due to an isotropic
energy dependent contamination from cosmic rays (CR).
A model for the CR background in diffuse class events
can be found in Figure 1 of Ref. \cite{2010PhRvL.104j1101A}.
The corresponding spectrum is rather hard, $\sim E^{-1}$,
reaching $\sim O(1)$ fraction of the total flux around 100 GeV
(compare Figure 1 and Table 1 in \cite{2010PhRvL.104j1101A}).

Gamma-ray flux with an index $n_{\rm d} \approx 2$ can be obtained by
inverse Compton scattering of interstellar radiation photons and a population of electrons 
with a spectrum $\sim E^{-3}$.
A break at a few hundred GeV can be explained by transition
between Thompson and Klein-Nishina scattering
for star-light photons, $E_{\rm break} \sim {(m_e c^2)^2}/{h \nu}$.
If we attribute the spherical signal with the ICS photons,
then the question would be to find a spherically symmetric source 
of high energy electrons at high latitudes.

\subsection{Limits on DM annihilation}
\lb{sect:dm_lim}

\begin{figure*}[th] 
\begin{center}
\hspace{-3mm}
\epsfig{figure = 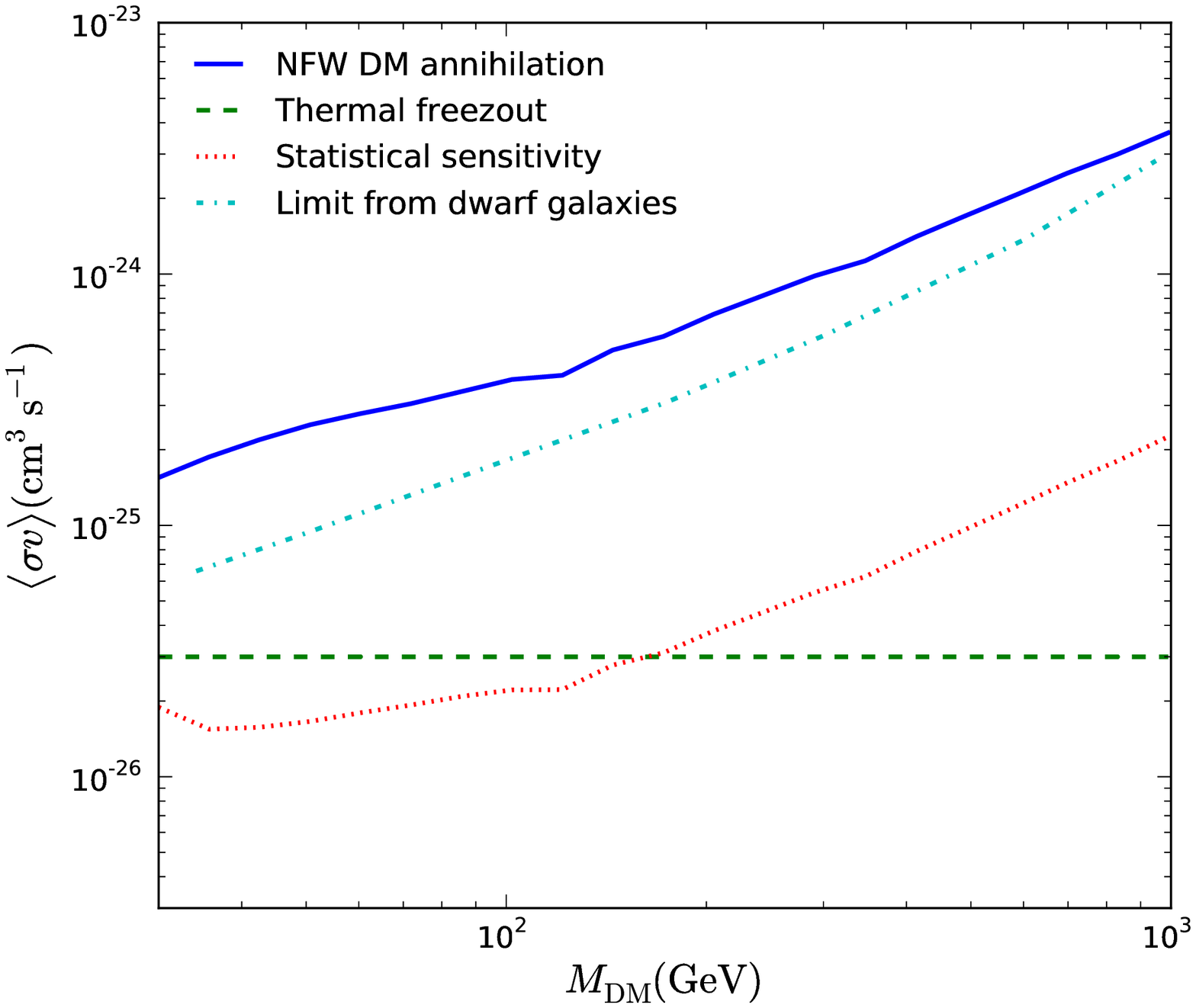,scale=0.42}
\epsfig{figure = 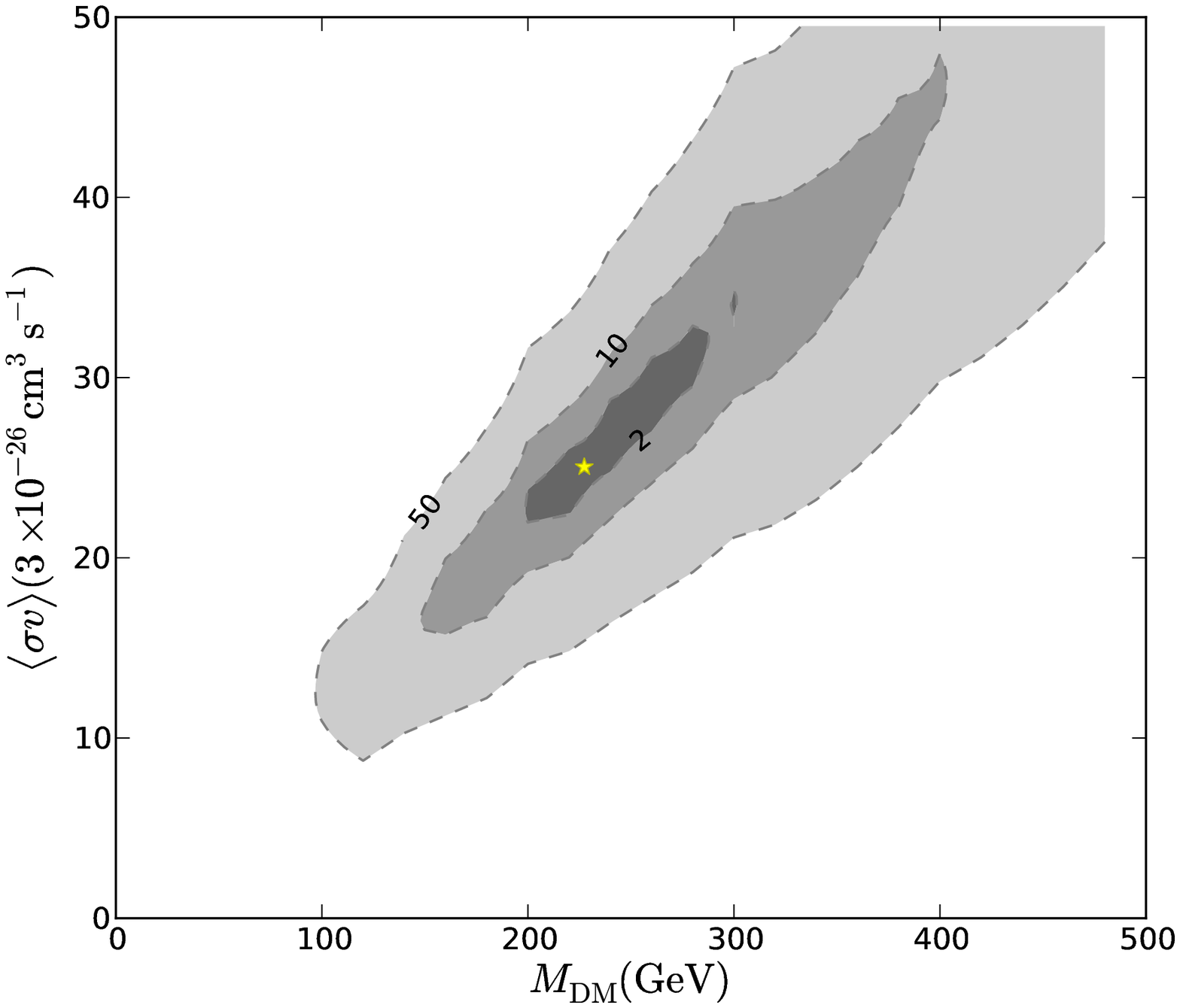,scale=0.42}
\end{center}
\vspace{-8mm}
\noindent
\caption{\small 
Left plot: upper bound on DM $b\bar{b}$ annihilation in an NFW profile
in comparison with the dwarf spheroidal galaxies limit \cite{2010ApJ...712..147A}.
The bound is 
derived by using the NFW annihilation flux in Figure \ref{fig:astroDM}
as an upper limit.
The statistical sensitivity is derived from the error bars for 
NFW flux in Figure \ref{fig:astroDM}.
Right plot: best $b\bar{b}$ DM annihilation fit to NFW flux points in 
Figure \ref{fig:astroDM}.
Contours represent $\Dl \chi^2$.
Limits from below are subject to systematic uncertainty due to modeling of astrophysical fluxes.
Limits from above may depend slightly on DM profile and annihilation channel.
}
\label{fig:DMlimits}
\vspace{1mm}
\end{figure*}

In this subsection we use 
the flux associated with the NFW annihilation template derived in Section \ref{sect:bin2bin}
to put conservative upper limits on the rate of DM annihilation
in the Milky Way halo.
Assuming $b\bar{b}$ annihilation channel, we find
the best-fit DM mass and annihilation cross section.
In the analysis, we use the prompt gamma-rays emitted
by the decay of $b\bar{b}$,
the corresponding spectrum is found with the help of the Pythia generator
\cite{2006JHEP...05..026S, 2008CoPhC.178..852S}.
The best-fit DM parameters are subject to significant systematic uncertainties
due to modeling of astrophysical emission.
The upper limits, on the other hand, are rather robust.
They only depend on the DM profile
and on the annihilation channel, e.g., $b\bar{b}$, $W^+W^-$, etc. 

Let $\bra \sm v \ket$ denote the DM annihilation cross section.
In this paper we will consider only the prompt gamma-ray emission 
from DM annihilation.
The flux of gamma-rays from DM annihilation 
per steradian at an angle $\theta$ from the GC is
\be
\lb{eq:DMline}
F_{\rm DM} (E, \theta) = \frac{1}{8\pi}  
\frac{\bra \sm v \ket}{M_{\rm DM}^2} \frac{dN_\g}{dE} \int \rho^2_{\rm DM}(r) dR
\ee
where $ \frac{dN_\g}{dE} $ is an average spectrum of prompt gamma-rays per annihilation event,
$R$ is the distance along the line-of-sight, and $r$ is the distance from GC,  
$r^2 = R^2 + R^2_0 + 2R R_0 \cos\theta$.
We assume local DM density
$\rho_{\rm DM{0}} = 0.4\:\text{GeV} \text{cm}^{-3}$
\cite{2009arXiv0907.0018C,2010A&A...509A..25W,2010arXiv1003.3101S}.

We parameterize the flux from DM annihilation by 
the DM mass and annihilation cross section.
The result of fitting the corresponding energy density of the flux to the best-fit fluxes
associated with the NFW annihilation template from Section \ref{sect:bin2bin}
is shown in Figure \ref{fig:DMlimits} on the right.
There are significant systematic uncertainties, 
e.g., a distribution of inverse Compton photons,
in proving the existence of a DM annihilation signal.
One can nevertheless put upper limits on DM annihilation,
provided that the flux from DM cannot be larger than the signal 
correlated with the NFW annihilation template.

In Figure \ref{fig:DMlimits} on the left for every DM mass $M_{\rm DM}$
we find the best-fit annihilation cross section which gives the upper limit
on DM annihilation.
Statistical uncertainty in this case is an order of magnitude smaller than the limit
itself.
This uncertainty provides an estimate of statistical sensitivity of our method 
for the DM annihilation signal.
For $M_{\rm DM} \lesssim 150$ GeV this sensitivity is sufficient
to detect DM annihilating with freeze-out cross section.

In spherical harmonics analysis the uncertainty is dominated by systematics
while statistical uncertainty is rather small.
This makes spherical harmonics a complimentary tool
to the searches of DM annihilation in dwarf spheroidal galaxies
\cite{2010ApJ...712..147A, 2011arXiv1102.5701L},
where the systematic uncertainties are small and the limits are dominated by statistics.

\section{\lb{sect:concl} Conclusions}

In this paper we argue that the spherical harmonics decomposition is a convenient tool to 
study the large-scale distribution of gamma-rays such as a possible
contribution from DM annihilation or decay.
The key points of this approach are the use of the spherical harmonics decomposition in a
window that eliminates most of the known astrophysical sources
and the choice of the coordinate system appropriate for the symmetries of 
DM distribution.
In Appendix \ref{sect:mct}
we show that in a test with $10^4$ randomly generated photons,
a 4\% fraction of gamma-rays coming from DM annihilation
can be detected with a five sigma significance.

One of the main advantages of the spherical harmonics decomposition 
compared to the analysis in coordinate space is an efficient organization of the data.
Consider, as an example, the top right plot in Figure 10.
The harmonics with $\ell \lesssim 20$ are dominated by the 
large-scale distribution of gamma-rays.
The harmonics between $\ell \sim 20$ and $\ell \sim 200$ are dominated
by the contribution from point sources,
while the harmonics above $\ell \sim 200$ are dominated by the noise
(either physical noise due to insufficient number of photons or the 
Point Spread Function (PSF) of the instrument).
Thus one can immediately
separate the data that carry some information about the large-scale distribution
of gamma-rays from the harmonics dominated by the noise.

Another approach in finding DM signatures in gamma-rays actively
discussed in the literature is to search for features in the power
spectrum due to DM subhalos 
\citep{2009PhRvD..80b3520A, 2009arXiv0912.1854H, 2010arXiv1005.0843C}.  
Our method is complimentary to this, since we
look for the signature of the main halo at small $\ell$, whereas DM
subhalos usually contribute at $\ell \gtrsim 100$.  We also believe that
with the current volume of data our approach is more advantageous
since $\ell \gtrsim 100$ harmonics are dominated by the noise and much
more data will be necessary to separate a significant signal, whereas
for small $\ell$ the current data is enough to overcome the noise level
for energies up to $30 - 50$ GeV.

Our method already enables us to argue that there is a significant
spherical distribution of photons in addition to an astrophysical flux,
which we model by taking the data in a low-energy bin as a template,
plus an isotropic distribution of photons.  
We compare several profiles with a spherical symmetry around the GC and 
find that below $\sim 50$ GeV ``stellar" halo $\propto 1/r^3$ 
has a slightly better $\chi^2$ than DM annihilation, DM decay,
or a bivariate Gaussian distribution found by \cite{2010ApJ...717..825D}. 

We also use the flux associated with the NFW annihilation template to
put upper limits on DM annihilation into $b\bar{b}$.
The derived limits are a factor of a few less stringent
than the limits from dwarf spheroidal galaxies.
The uncertainty of our method is dominated by systematics
while the dwarf spheroidal galaxies (dSph) method is dominated by statistics.
One of the advantages of using the DM annihilation in the Milky Way halo
versus the annihilation in dSph is the ability to use the \Fermi data
to put stronger constraints on DM annihilation,
even after the \Fermi LAT stops collecting data,
by reducing the systematic uncertainty 
related to modeling the astrophysics backgrounds.

\medskip

{\large \bf Acknowledgments.}

\noindent
The authors are thankful to Douglas Finkbeiner, Jennifer Siegal-Gaskins, 
Neal Weiner, and especially to David Hogg for
valuable discussions and comments.  
This work is supported in part by the Russian Foundation of Basic Research under 
Grant No. RFBR 10-02-01315 (D.M.), by the NSF Grant No. PHY-0758032 (D.M.), 
by DOE OJI Grant No. DE-FG02-06E R41417 (I.C.), 
by the Mark Leslie Graduate Assistantship NYU (I.C.), 
by NASA Grant NNX08AJ48G (J.B.), and by the NSF Grant AST-0908357 (J.B.). 
Some of the results in this paper have been
derived using the HEALPix \citep{2005ApJ...622..759G} package.

\appendix

\section{Spherical harmonics covariance}
\lb{sect:shcov}

In this appendix we provide details on the calculation of covariance matrix
for spherical harmonics defined in a window on the sphere.

Consider a pixelation of the sphere.
We will assume that the pixel size is sufficiently small compared to the angular size of interest,
but it is sufficiently large so that the photons in different pixels are uncorrelated.
Denote by $n_p$ the number of photons in a pixel $p$,
let $\g_p$ be the center of $p$, and $\dl \Om_p$ be the area of $p$.
We can define a discrete photon density function $\rho_p = \frac{n_p}{\dl \Om_p}$.
The spherical harmonics transform of the density function is
\be
a_{lm} = \int Y^*_{lm}(\g) \rho(\g) d\Om
\approx \sum_{p = 1}^{N_{\rm pix}} Y^*_{lm}(\g_p) n_p,
\ee
The spherical functions are defined as
\be
Y_{lm} (\theta,\: \vp) = \sqrt{\frac{2l+1}{4\pi} \frac{(l - m)!}{(l + m)!}} 
P_{l}^{m} (\cos\theta) e^{i m \vp}\,.
\ee 
The spherical harmonics are orthogonal 
on the sphere but on a window in the sphere they are not.
As a result, we expect correlation between different harmonics.

The covariance matrix of $ a_{lm}$'s is
\bea
\lb{eq:alm_vm}
C_{lm,\, l'm'} &=& \bra  a_{lm}^*  a_{l'm'} \ket - \bra  a_{lm}^*\ket \bra  a_{l'm'} \ket  \\
\nonumber
&=& \sum_p Y^*_{l'm'}(\g_p) Y_{lm}(\g_p) (\bra   n_p^2 \ket - \bra   n_p \ket^2),
\eea
where we have used that the numbers of photons at different points are not correlated 
$\bra n_p n_{p'} \ket = \bra n_p \ket \bra n_{p'} \ket$.

For a Poisson distribution $\bra   n_p^2 \ket - \bra n_p \ket^2 = \bra n_p \ket$.
In a particular realization of the photon map,
our best estimate of $\bra n_p \ket$ is the actual number of photons 
in this pixel $n_p$.
Consequently, the covariance matrix can be estimated as
\be
C_{lm,\, l'm'} = \sum_p Y^*_{l'm'}(\g_p) Y_{lm}(\g_p) n_p\,.
\ee
Spherical harmonics coefficients $a_{lm}$ together with their covariance matrix
provide the data necessary to formulate 
a $\chi^2$ fitting procedure.
Denote by $b_{lm}(\al)$ the spherical harmonics decomposition of a model
prediction for the distribution of gamma-rays depending
on a set $\al$ of parameters describing the model.
The best-fit parameters can be found by minimizing
\be
\lb{eq:chi2_ini}
\chi^2(\al) = \sum_{lm,\, l'm'} C_{lm,\, l'm'}^{-1} (a_{lm}^* - b_{lm}^*(\al)) (a_{l'm'} - b_{l'm'}(\al)),
\ee
where the star denotes complex conjugate and $C_{lm,\, l'm'}^{-1}$ is the inverse of the covariance matrix
\be
\sum_{l'm'} C_{lm,\, l'm'}^{-1} C_{l'm',\, l''m''} = \dl_{l l''} \dl_{m m''}.
\ee


\section{\lb{sect:mct} Monte Carlo test}

\begin{figure}[t] 
\begin{center}
\epsfig{figure = 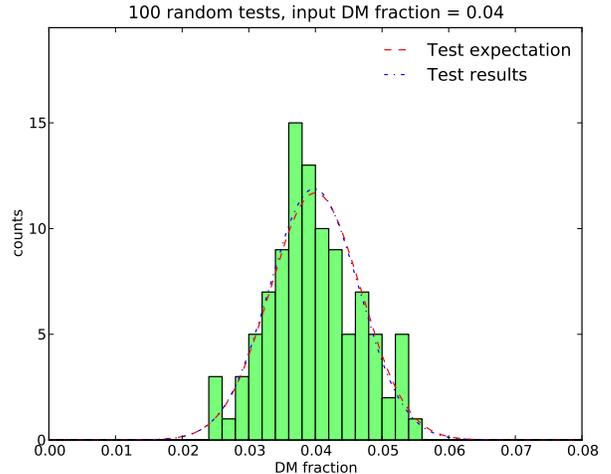, scale=.45}
\end{center}
\vspace{-8mm}
\noindent
\caption{\small 
Random test of the spherical harmonics decomposition algorithm for separating a 
DM NFW annihilation signal from an isotropic background.
The input DM fraction is 0.04.
Plotted fractions represent the result of fitting in spherical harmonics space.
An approximation of these fractions by a Gaussian distribution is the ``Test results'' curve.
``Test expectation'' is a Gaussian around 0.04 with the scatter derived from
Equation (\ref{eq:var_f}).
}
\label{fig:rndTest}
\vspace{1mm}
\end{figure}

In this appendix we check our method for separating a DM fraction
by generating random distributions of photons.
For the test we use an isotropic distribution 
plus a distribution coming from DM annihilation in an NFW profile
(Equation (\ref{eq:NFW})).

The model parameters are the same as in Section \ref{sect:data}:
the window is $|b| > 10^\circ$;
we use $Y_{lm}$ harmonics with $l = 0,\ldots,15$ and $m = 0$;
the HEALPix parameter is nside = 32.

The average total number of photons inside the window
is $\nphrnd$ with an average fraction of photons coming from
DM annihilation \dmFracRND.
In every pixel $p$ we put a random number of photons $n_p$
according to the Poisson distribution with an average
equal to the combined density at that pixel $\mu_p = \mu_p^{\rm isotr} + \mu_p^{\rm DM}$.

In the test we generated $N_{\rm tests} = \ntests$ realizations of the photon map.
For every realization $i$, we find the best estimate of the DM fraction
$q^{\rm DM}_i$.
The average among the realization and the standard deviation are 
$(3.98 \pm 0.67)\times 10^{-2}$.
The corresponding distribution is presented in Figure \ref{fig:rndTest}
as ``Test results.''

In real applications, there is usually only one realization of the data available,
i.e., we need a way to estimate the uncertainty of the result
based only on one realization.
This uncertainty can be estimated from 
the curvature of $\chi^2$ near the minimum in the direction of the parameter
(Equation (\ref{eq:var_f})).
The uncertainty derived from $\chi^2$ (averaged over the realizations) is
$\sm =  0.68 \times 10^{-2}$.
The corresponding Gaussian distribution
$(4.00 \pm 0.68)\times 10^{-2}$
is plotted in Figure \ref{fig:rndTest} as ``Test expectation.''

The expected deviation of the mean is 
$\sigma / \sqrt{N_{\rm tests}} \approx 0.07$.
Thus the actual deviation of the mean DM fraction
from the expected value is less than one sigma.
Also the difference of the actual standard deviation and the expected one
is less than one sigma.
We conclude that, given a particular distribution of photons, the best estimate of the variance 
given by Equation (\ref{eq:var_f})
is an adequate representation of the actual variance among the realizations 
of the photon distribution.

The  $\chi^2 = 13.9 \pm 5.2$ is calculated from Equation (\ref{eq:chi2_ini}).
The number of degrees of freedom is fourteen:
there are sixteen data points corresponding to $Y_{l0}$ harmonics for
$l = 0,\ldots, \lmax$
and two varying parameters corresponding to
the normalization of the two templates: isotropic and DM.

We find that the spherical harmonics decomposition
is a statistically unbiased fitting method
with a viable estimation of statistical uncertainty.

\section{\lb{sect:ps} Point sources}

\begin{figure*}[h] 
\begin{center}
\epsfig{figure = 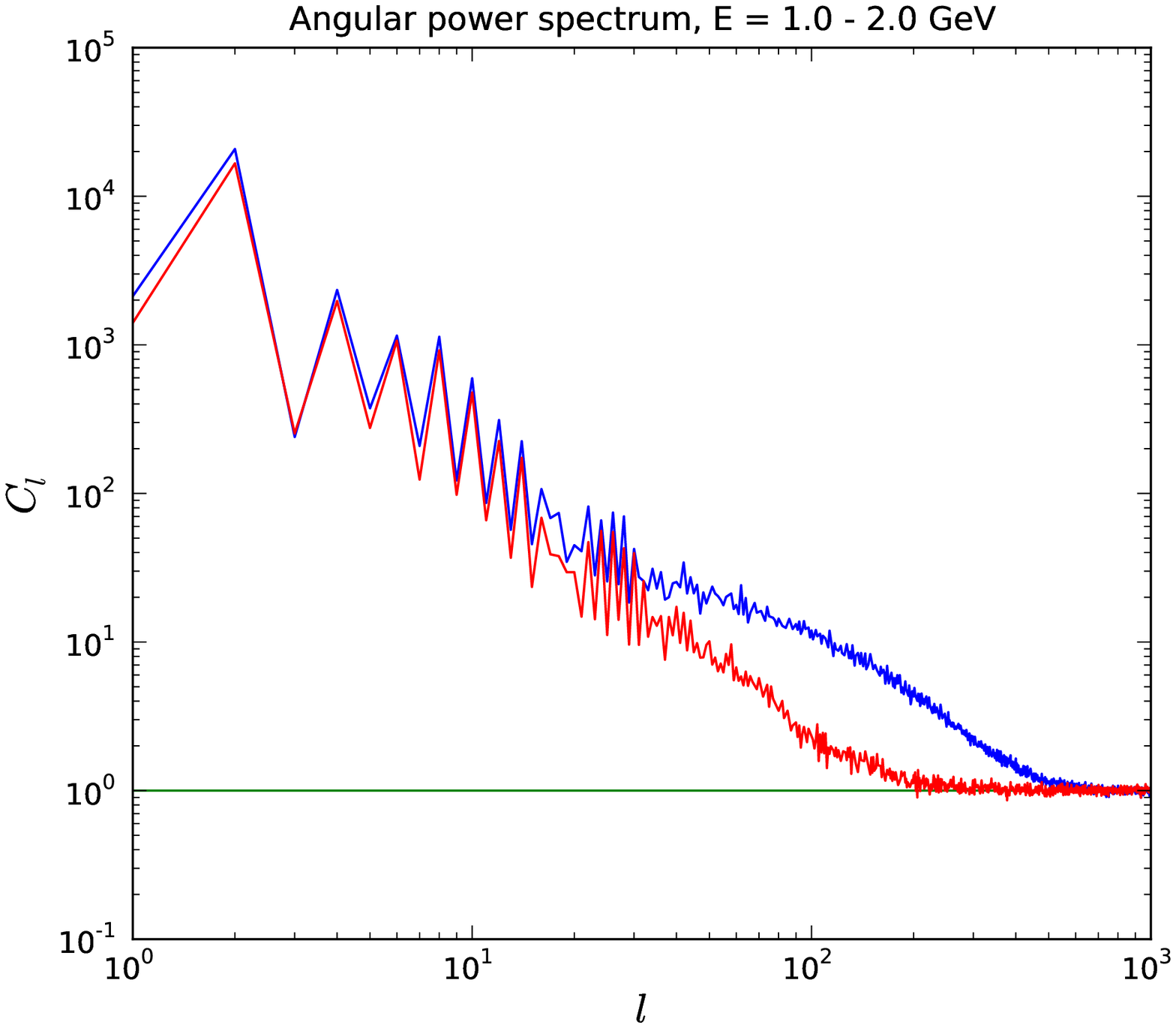,scale=\smallScale}
\epsfig{figure = 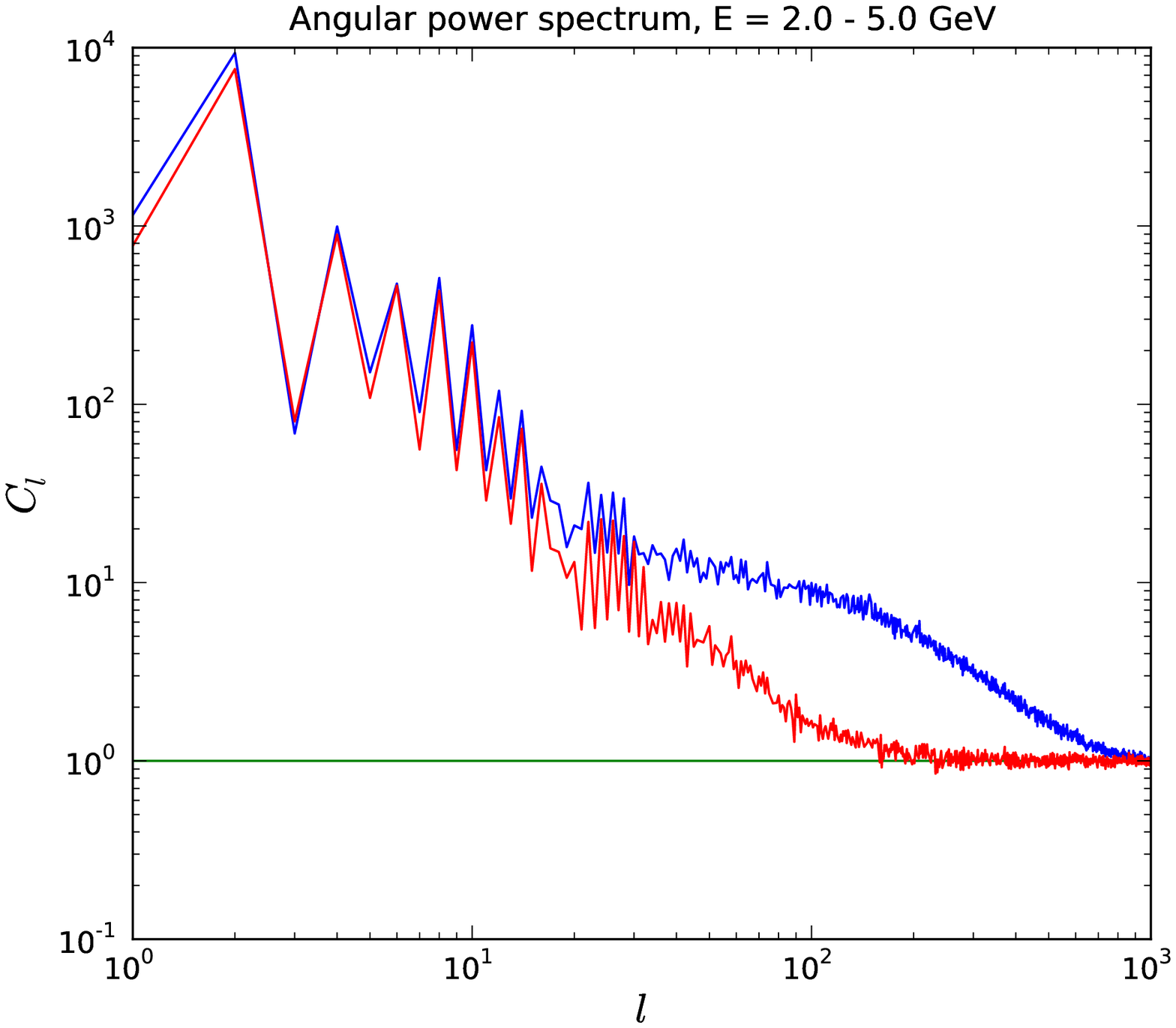,scale=\smallScale}
\epsfig{figure = 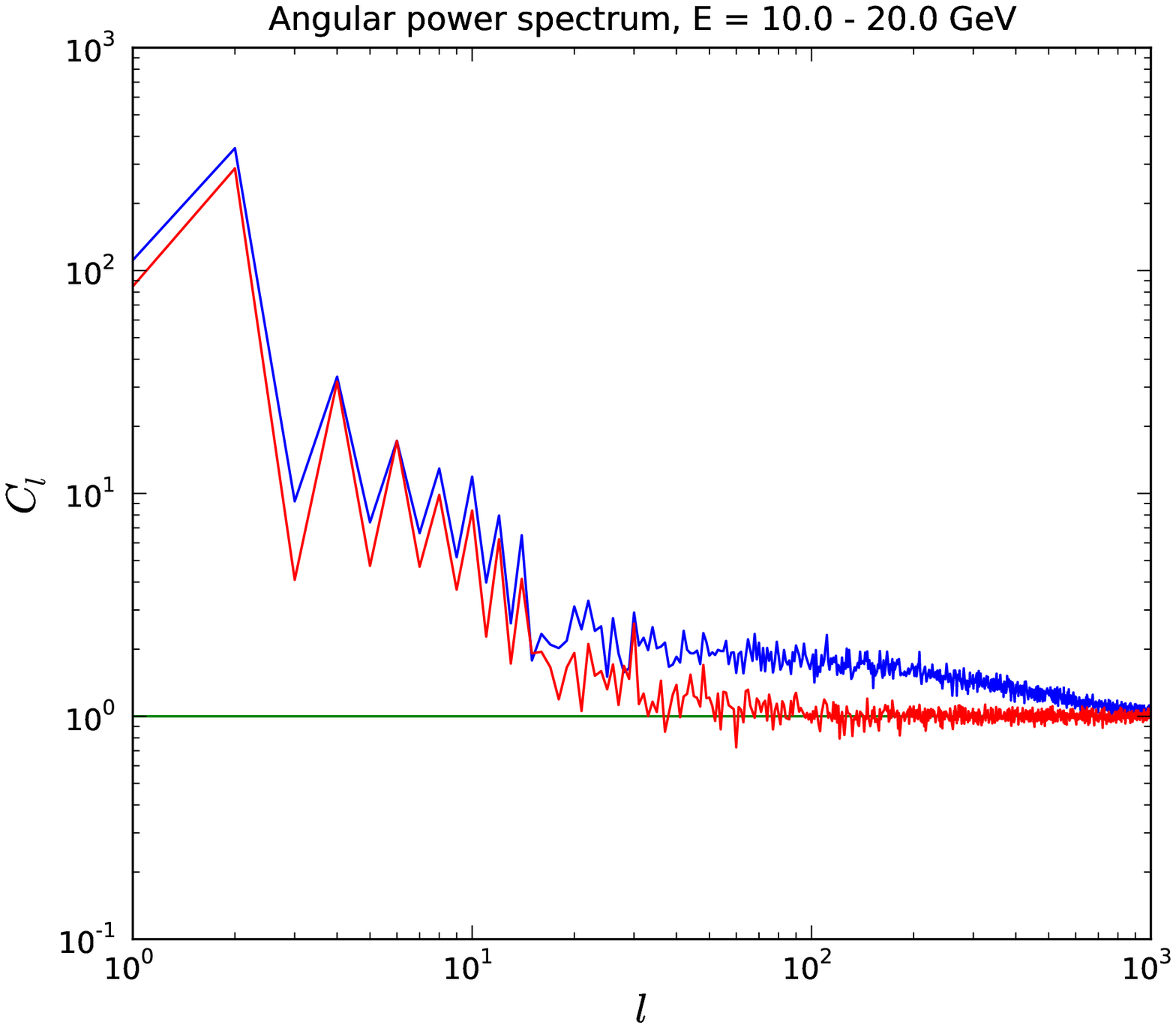,scale=\smallScale}
\epsfig{figure = 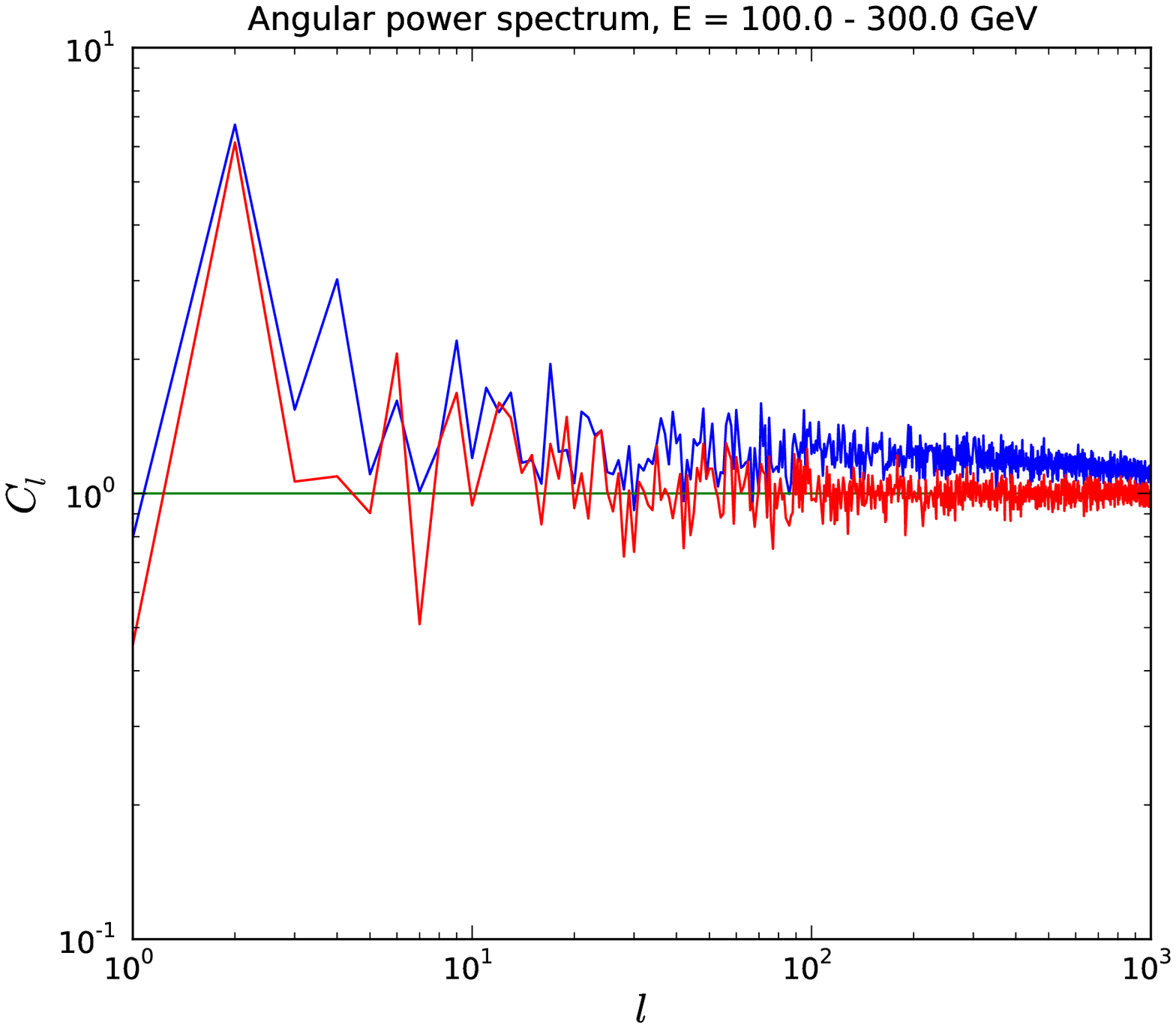,scale=\smallScale}
\end{center}
\vspace{-8mm}
\noindent
\caption{\small 
Power spectrum of angular modes for \Fermi gamma-ray data inside a window
excluding the Galactic plane $|b|>10^\circ$ for some characteristic energy bins.
Blue (upper) lines: all gamma-rays are taken into account, including point sources.
Red (lower) lines: \Fermi gamma-ray point sources \cite{2010ApJS..188..405A}
are masked.
The normalization is chosen such that  $\bra C_l \ket = 1$ 
for the Poisson noise (green constant lines).
At low $\ell$ the power spectrum is dominated by the large-scale structure
(the red and the blue curves are almost identical).
At $\ell \sim 100$ the $C_l$'s are dominated by the point sources
(the blue curve is significantly higher than the red curve).
For zero PSF, the blue curve would stay for large $\ell$ at a constant level,
given by Equation (C11), above the Poisson noise level.
The suppression of $C_l$'s to the Poisson noise level for higher $\ell$ 
is due to non-zero PSF.
In the highest energy bin, 100 - 300 GeV, 
the PSF is less than $0.1^\circ$ \cite{2009ApJ...697.1071A}
which corresponds to $\ell > 2000$.
As a result, the blue curve stays above the red curve in this energy bin.
At lower energies the PSF is higher and the suppression of the angular
power spectrum to the Poisson noise level occurs for smaller $\ell$'s.
In the plots, we use the HEALPix parameter nside = 1024.
}
\label{fig:mav}
\vspace{1mm}
\end{figure*}

In this appendix we study the dependence of spherical harmonics on the contribution from point sources.
We find that in the presence of point sources the variance of $a_{lm}$'s increases.

We will assume some pixelation of the sphere with pixels of equal area.
Denote by $n_p$ the number of photons in a pixel $p$ and
define the spherical harmonics coefficients
\be
\lb{eq:alm_norm}
a_{lm} = \sqrt{\frac{4\pi}{N_\g}} \sum_p Y^*_{lm}(\g_p) n_p
\ee
where the sum is over the pixels, $\g_p$ is the center of pixel $p$, and
$N_\g$ is the total number of photons.
The normalization of $a_{lm}$'s in this appendix is different 
from the normalization everywhere else in the paper.
We show below that with this normalization
the expected variance of spherical harmonics in the case of Poisson noise
is equal to one.

Analogously to the derivation of the covariance matrix in Equation (\ref{eq:alm_vm}),
we find that the variance of $a_{lm}$'s is
\be
\lb{eq:alm_var}
{\rm Var} (a_{lm}) = \frac{4\pi}{N_\g} \sum_p Y^*_{lm}(\g_p) Y_{lm}(\g_p) {\rm Var} (n_p)
\ee
If we assume that the diffuse emission and the point sources were distributed isotropically,
then $\bra a_{lm} \ket = 0$ and 
 $\bra a_{lm}^2 \ket = \bra a_{lm'}^2 \ket$ for any $l,\, m, \, m'$. 
In this case, one can relate the variance of spherical harmonics to the expectation value
of the angular power spectrum
\be
\lb{eq:varalm_cl}
{\rm Var} (a_{lm}) = \frac{1}{2l + 1} \sum_{m = -l}^l {\rm Var} (a_{lm}) = \bra C_l \ket.
\ee
In the following we will show that this is a good approximation for large $\ell$,
whereas small $\ell$ harmonics are dominated by the 
non-isotropic Galactic emission
with $\bra a_{lm} \ket \neq 0$.

In the case of the Poisson statistics, the best estimate for the variance 
of the number of photons is ${\rm Var} (n_p) = n_p$.
Taking into account that for any point on the sphere
\be
\lb{eq:sf_s2}
 \frac{1}{2l + 1} \sum_m Y^*_{lm}(\g) Y_{lm}(\g) = \frac{1}{4\pi}
\ee
we find from Equations (\ref{eq:alm_var} - \ref{eq:sf_s2}) that
in the case of the Poisson noise
\be
\lb{eq:var_poisson}
{\rm Var} (a_{lm}) 
= \frac{1}{N_\g} \sum_p {\rm Var} (n_p)
= 1
\ee
Now suppose that there are some point sources.
Denote by $x_m$ the expected number of $m$-photon sources inside a pixel.
In this case the statistics of photons in pixels across the sky is not Poisson.
Instead, we will assume the Poisson statistics of the point sources
with the average values $x_m$.
In particular the variance of the number of photon sources is 
\be
{\rm Var}(x_m) = x_m
\ee
The variance of the number of photons in a pixel
is the sum of the variances of the photon sources
times the number of photons from every source squared
\be
{\rm Var}(n_p) = \sum_{m} m^2 {\rm Var}(x_m) = \sum_{m} m^2 x_m
\ee
Let us introduce the following parameter
\be
\lb{eq:mavdef}
m_{\rm av} = \frac{\sum_m m^2  x_m}{\sum_m m  x_m}\,,
\ee
where ${\sum_m m  x_m} = \bra n_p \ket$ is the expected number 
of photons in a pixel.
If there is a significant
contribution of multi-photon point sources to gamma-ray data,
then $m_{\rm av} \gg 1$, 
while $m_{\rm av} = 1$ for truly diffuse emission.
In analogy with Equation (\ref{eq:var_poisson}), we find
\be
{\rm Var} (a_{lm}) 
= \frac{1}{N_\g} \sum_p {\rm Var} (n_p)
= m_{\rm av}
\ee

To summarize,
for isotropic distribution of photons and
for the angular scales smaller than the detector PSF,
we expect 
\be
\bra C_l \ket = 1.
\ee
This limit should be saturated for sufficiently large $\ell$.

In the presence of point sources, the variance is $m_{\rm av}$ times
larger than the variance in the Poisson statistics case.
Consequently,
for isotropic distribution of point sources
(or when the angular scale corresponding to $\ell$ is much smaller
than the scale of the distribution), we expect
\be
\lb{eq:pscl}
\bra C_l \ket = m_{\rm av}\, .
\ee
We expect this behavior for intermediate values of $\ell$.
At small $\ell$, the $C_l$'s are dominated by a large-scale
distribution of gamma-rays.

In Figure 10 we compare the angular power spectra 
before and after masking the gamma-ray point sources
for several characteristic energy bins.
We use the same \Fermi data as described in Section
\ref{sect:data} and mask the Galactic plane within $|b| < 10^\circ$.
For $\ell \lesssim 20$ the angular power spectrum is dominated
by the large-scale distribution of photons:
these values of $\ell$ are of interest for fitting templates at large angular scales
using the spherical harmonics.
For intermediate $\ell$'s the angular power spectrum is dominated by the contribution
from point sources.
For large $\ell$ the angular power spectrum is consistent with the Poisson
noise with an exception of the highest energy bin (100 - 300 GeV),
where the signal in the presence of point sources is above the Poisson noise level
even for $\ell \sim 1000$ corresponding to angular scales $\sim 0.2^\circ$.
This is consistent with the 
\href{http://www-glast.slac.stanford.edu/software/IS/glast_lat_performance.htm}{PSF}
$\lesssim 0.1^\circ$ for gamma-rays with energies
above 100 GeV \cite{2009ApJ...697.1071A}.

In the analysis we use the HEALPix parameter
nside = 1024 which corresponds to approximately $10^7$ pixels.
The ``pixelized" values of $C_l$'s are supposed to be smaller than the
values computed by a continuous integration, $C_{l\: \rm pix} = w_l^2 C_{l\: \rm cont}$,
where $w_l$ is the pixel window function.
The window function is a decreasing function which is equal to 1 at $\ell = 0$.
For nside = 1024 and $\ell = 1000$, $w_l^2 \approx 0.956$,
i.e. the values of $C_l$'s in Figure 10 are less than about
5\% smaller than the real values.
For nside = 32 the window function at $\ell = 15$ is $w_l = 0.989$.
We did not rescale the spherical harmonics by the window function
in the fitting procedure.
The only effect of the window function is to put a little more weight on
lower harmonics.

Let us make 
one more technical comment about the derivation of the plots in
Figure 10.
In order to relate the variance in the spherical harmonics to the value of $C_l$'s
we need the average $\bra a_{lm} \ket = 0$.
In the analysis 
we have used the spherical harmonics decomposition in a window
$|b| > 10^\circ$
(the window is the same as defined in section \ref{sect:data}).
For the spherical harmonics decomposition in a window,
at least some of the expected 
$a_{lm}$'s are nonzero.
In order to make them zero without affecting the variance we subtract
the spherical harmonics of an isotropic distribution inside the window.



\bibliography{HFpapers}         

\end{document}